\magnification=1200
\parskip=10pt plus 5pt
\parindent=15pt
\baselineskip=15pt
\input amssym.def
\input amssym
\topglue 0.6in
\pageno=0
\footline={\ifnum \pageno < 1 \else \hss \folio \hss \fi}
\line{\hfil{August, 2000}}
\vskip .8in
\centerline{\bf THE EFFECT OF HIGHER-ORDER CURVATURE TERMS}
\centerline{\bf ON STRING QUANTUM COSMOLOGY}
\vskip .5in
\centerline{Simon Davis and Hugh Luckock}
\vskip 20pt
\centerline{School of Mathematics and Statistics}
\centerline{University of Sydney}
\centerline{NSW 2006}
\vskip 20pt
\centerline{\bf Abstract}  
Several new results regarding the quantum cosmology of higher-order
gravity theories derived from superstring effective actions are
presented.  After describing techniques for solving the Wheeler-DeWitt 
equation with appropriate boundary conditions, it is shown that
this quantum cosmological model may be compared with semiclassical 
theories of inflationary cosmology.  In particular, it should be possible
to compute corrections to the standard inflationary model perturbatively   
about a stable exponentially expanding classical background.

\vfill\eject

\noindent
{\bf 1. Classical Cosmological Space-times of Higher-Order Gravity Theories}

Since solutions to the general relativistic field equations 
contain initial curvature singularities whenever the dominant
energy condition is satisfied, one of the motivations for developing
quantum cosmology has been the theoretical justification of the elimination
of the singularities.  Curvature singularities predicted by general 
relativity can be avoided by introducing boundary conditions in the path 
integral defining the quantum theory which restrict the integral over all 
four-geometries and matter fields to Riemannian metrics $g_{\mu\nu}$ 
on compact manifolds bounded by a three-dimensional hypersurface with metric 
$h_{ij}$ and fields with specified values on the hypersurface [1].  
They might also be eliminated in a quantum theory of gravity free from 
divergences at short distances, since the wavefunctions defined by the path 
integral of the theory may represent non-singular geometries at initial 
times.

Classical cosmological solutions to the equations of motion 
for several different types of theories containing higher-order curvature 
terms  have been analyzed with regard also to the presence 
of singularities, and non-singular solutions have been obtained.
Renormalizability has been obtained with the addition of quadratic 
terms in the action [2].  Moreover, any theory of superstrings consistent 
at the quantum level will have an effective action containing 
higher-order curvature terms.  It has been shown, in particular, that 
dimensionally continued Lovelock invariants, giving rise to second-order
field equations for the metric, arise in superstring effective actions.  

According to the conventional usage of terms related to modifications 
of the Einstein-Hilbert action, a Lagrangian which has the form 
$L(g_{\mu\nu},\partial_\alpha g_{\mu\nu}, \partial_\alpha 
\partial_\beta g_{\mu\nu})$, and is a nonlinear function of the curvature 
tensor, represents a higher-order gravity theory [3], whereas an action 
containing third- and higher-order derivatives of the metric describes 
higher-derivative gravity theories.    

To determine whether quantum cosmological models based on theories
containing third- and higher-order derivatives of the metric or powers of
the second derivative of the metric produce wave functions consistent with a 
non-singular geometry, without imposing a non-singular boundary condition at 
initial times, it is useful to start with an action which combines 
higher-order gravity with scalar fields and possesses singularity-free 
cosmological solutions [4][5].
At string tree-level and first-order in the $\alpha^\prime$-expansion
of the compactified heterotic string effective action in four dimensions,
the dynamics of the graviton, scalar field S and modulus field T can be
described by the effective Lagrangian
$${\cal L}_{eff.}~=~{1\over {2\kappa^2}}R+
{{DS D{\bar S}}\over {(S+{\bar S})^2}}
+3{{DT D{\bar T}}\over {(T+{\bar T})^2}}
+{1\over 8}(Re~S)R_{GB}^2+{1\over 8}(Im~S)R {\tilde R}
-{1\over 2}V(S,{\bar S},T,{\bar T})
\eqno(1)
$$
If $Re~T$, representing the square of the compactification radius, is
set equal to a constant, and $Im~T$ is set equal to zero, the kinetic 
term for the modulus field vanishes.  In addition, defining the 
real part of the dilaton field to be $Re~S~=~{1\over {g_4^2}}~e^{-\Phi}$
restricting attention to isotropic models for which $R{\tilde R}=0$ , and 
choosing units such that $\kappa~=~1$, one obtains the action 
\footnote*{The sign convention in reference [6] is being used for the 
definition of $Re~S$ and it is combined with the expression 
for ${\cal L}_{eff.}$ in reference [4].  The integral for the action $I$ is 
deduced after multiplication by an overall factor of 2.}  
$$I~=~\int~d^4x~{\sqrt{-g}}~\left[~R~+~{1\over 2}(D\Phi)^2
~+~{{e^{-\Phi}}\over {4g_4^2}}(R_{\mu\nu\kappa\lambda} 
R^{\mu\nu\kappa\lambda} - 4 R_{\mu\nu} R^{\mu\nu} + R^2)~-~V(\Phi)\right]
\eqno(2)
$$
The Gauss-Bonnet invariant arises in this action,
but it is multiplied now by the factor
${{e^{-\Phi}}\over {4 g_4^2}}$, where $g_4$ is the four-dimensional
string coupling constant and $\Phi$ is the dilaton field, so that the
integral is not a topological invariant.  This quadratic curvature 
combination is precisely that required to remove ghost poles in 
the perturbative expansion of the propagator, and since the term is dynamical, 
the theory represents a modification of general relativity which is unitary 
and has improved renormalizability properties.

There are conformal transformations which map $f(R)$ actions to 
the Einstein-Hilbert action plus a scalar field with a potential term [7]-[13].
Other field redefinitions can be used to transform $f(R_{\mu\nu}
R^{\mu\nu})$ to the Ricci scalar plus terms containing scalar fields and 
extra tensor modes [14]-[16].  The transformed theories should have the same 
renormalizability properties as the higher-order gravity theories, 
because field redefinitions do not alter the S-matrix.  Such transformations
are unnecessary in the present case since the field equations obtained
from equation (2) contain at most two derivatives - in effect, although the
action contains higher-order curvature terms, the addition of a boundary term
is sufficient to eliminate second-order derivatives of the metric from the
integral, so that only a conventional quantization of the theory is required, 
rather than the Ostrogradski method [3]. 

It is known that most classical string equations of motion without a 
dilaton field do not lead to inflation [17].  Thus, dilaton fields also have 
been included as they permit inflation, and a set of higher-order gravity 
theories with a dilaton field has been shown to produce the 
required inflationary growth of the Friedmann-Robertson-Walker scale 
factor [18].  

An earlier analysis of $R^2$ theories [19] and $C^2$
theories [20] has shown that an $R^2$ term leads to particle production and 
inflation with minimal dependence on the initial conditions, while  
the $C^2$ term  generates large anisotropy [21] and causes destabilization of
positive $\Lambda$ metrics.  Inflation has been also derived from 
higher-derivative terms directly obtained as renormalization 
counterterms [22][23] without the inclusion of scalar or inflaton fields.

\noindent{\bf 2. The Quantum Cosmology of Higher-Order Gravity Theories}
\vskip 2pt

Much of the initial work on higher-derivative and higher-order
quantum cosmology has been developed with only curvature terms and 
no scalar field in the action.  Nevertheless, scalar fields are probably 
necessary in a theory describing
gravity and matter interactions, because they are required for 
renormalizability of gauge theories with massive spin-1 vector fields [24] 
and they result from the conformal rescaling of the metric 
in higher-order theories [7]-[13].

The quantum cosmology of standard gravity coupled to a scalar field has 
been investigated by many authors [25]-[28].  The non-zero vacuum
expectation value of the scalar field in grand unified theories drives 
inflation in semi-classical cosmology, and again it is found to be useful 
in obtaining wave functions representing inflationary solutions in 
quantum cosmology.   

The quantum cosmology of superstring and heterotic string effective 
actions in ten dimensions with curvature terms up to fourth order 
has been investigated previously [29]-[32]. With the dilaton
and modulus fields included, the Wheeler-DeWitt equations for both theories,
in the minisuperspace of metrics with different scale factors for the
physical and internal spaces, have not been solved in closed form, because
the coupling of the curvature and scalar fields leads to 
quartic terms in the derivatives of the scale factor in the Hamiltonian
and fractional powers of the momenta [30].  For the superstring, the 
differential is only reducible to the form of a 
diffusion equation when the curvatures of the physical and internal 
spaces are set equal to zero, and the scalar field is set equal 
to a constant [32].  

When there are quadratic curvature terms in the heterotic string effective 
action, the Hamiltonian cannot be expressed as a simple function of the 
canonical momenta, preventing a derivation of a directly solvable
Wheeler-DeWitt equation.  For the model in this paper, a pseudo-differential 
equation in closed form, obtained by using the Ostrogradski 
formalism for the heterotic string effective action (2), can be converted 
to a partial differential equation of higher order, which may be solved 
with appropriate boundary conditions [33]. 

The Hamiltonian is more directly obtained if one begins by adding a total 
derivative term to the Lagrangian to eliminate second-order derivatives 
and a different set of momenta conjugate to the scale factor and dilaton 
field only is used.  While the first choice of momenta gives rise to a 
sixth-order Wheeler-DeWitt equation, the second set of momenta 
results in a Wheeler-DeWitt equation which reproduces the 
standard equation for gravity coupled to a scalar field in the limit that 
the quadratic term vanishes.  The wave function can then be determined 
perturbatively in the dilaton coupling ${{e^{-\Phi}}\over {g_4^2}}$.  

Similarly, there is a preference for quantizing 
a theory that does not include higher-order curvature terms, because the 
absence of higher derivatives of the metric allows one to avoid
superfluous degrees of freedom.  However, if the required field redefinitions
depend on the Riemann curvature tensor and its contractions, it is the set 
of kinetic terms for the extra scalar fields which now introduce the 
higher derivatives of the metric, implying that additional 
difficulties arising in quantization would not necessarily be circumvented 
for these types of higher-order theories. 
 
While the coupling ${{e^{-\Phi}}\over {g_4^2}}$ explicitly determines the 
strength of higher-order terms relative to the Ricci scalar, a factor of 
$\hbar$ can be restored at each order in the perturbative expansion of the 
S-matrix of the heterotic string sigma model.  Since the $(n+1)-th$ loop 
computation for the sigma model produces the $n^{th}$-order contribution to 
the heterotic string effective action $I_{eff.}^{(n)}$, 
the limit $\hbar \to 0$ can be taken, after multiplication by a factor of 
${1\over \hbar}$, to obtain the classical action for gravity coupled 
to a scalar field.  It is necessary to take the 
$\hbar \to 0$ limit because the ${{e^{-\Phi}}\over
{g_4^2}} \to 0$ limit eliminates the dilaton field kinetic term. 
Nevertheless, since additional factors of $\hbar$ and 
${{e^{-\Phi}}\over {g_4^2}}$ occur simultaneously at higher loops in the 
sigma model perturbation expansion, the Wheeler-DeWitt 
equation for the higher-order gravity theory 
can be regarded as a perturbation of the standard second-order 
Wheeler-DeWitt equation, with the solution being expanded in 
powers of ${{e^{-\Phi}}\over {g_4^2}}$.  This provides an approximate wave 
function for the heterotic string effective action which includes 
corrections to the wave function used to predict inflationary cosmology.  

Given a fundamental theory at Planck scale with higher-order curvature terms, 
it is appropriate to consider a boundary located between the Planck era and
the inflationary epoch where the predictions of quantum cosmology of the
higher-derivative theory could be matched, in principle, to the predictions
of the quantum theory of standard gravity coupled to matter fields.
The inclusion of this boundary will have an effect on both the
quantum cosmology of the more fundamental theory and the computation
of radiative corrections to the semi-classical inflationary model.

The significance of the wave function depends on the stability 
of the classical background geometries which represent most probable 
configurations in the quantum cosmology of the model.  Stability at the 
nonlinear level can be proven if the positive energy theorem holds, which 
requires that the space-time admits a Killing spinor that also must be a 
supersymmetry parameter.  For the four-dimensional compactification 
of heterotic string theory, the Majorana condition on the Killing 
spinor leads to differential equation for the scale factor which is 
solved by a non-singular cosmological bounce solution.  When this scale
factor is substituted into the equation of motion for the dilaton, it
will be shown that there is a solution for $\Phi$ which increases at an 
approximately linear rate with respect to time at large $t$.  Thus, there is
a stable background configuration derived from the heterotic string effective
action, which describes the exponential expansion of the inflationary
epoch and allows for the computation of perturbative corrections in the
quantum cosmological model.

\vskip 10pt
\noindent
{\bf 3. Quantum Cosmology for a Four-Dimensional Heterotic String Effective}
\hfil\break
\phantom{....}{\bf Action}

The model (2) containing quadratic curvature terms and the dilaton field
can be quantized with the Hamiltonian constraint being represented by
the Wheeler-DeWitt equation.  Since the solution to this equation generally
requires a reduction in the number of degrees of freedom in the metric
field, it is convenient to consider only a minisuperspace
of Friedmann-Robertson-Walker metrics 
$g_{\mu\nu}=diag\left(1,-{{a^2(t)}\over {1-Kr^2}},-a^2(t)r^2,-a^2(t)
r^2 sin~\theta \right)$ with $K=1$ (closed model), 
$K=0$ (spatially flat model) or $K=-1$ (open model). 

Homogeneity of the minisuperspace model implies that the fields are 
position-
\hfil\break
independent on foliations of the four-dimensional space-times
and the action per unit volume is a one-dimensional integral
$${I\over {\bar{\cal V}}}~=~
\int~dt~\biggl[(6a^2 {\ddot a}~+~6a{\dot a}^2~+~6aK)~+~{1\over 2}
a^3~{\dot\Phi}^2~+~6{{e^{-\Phi}}\over {g_4^2}}{\ddot a}({\dot a}^2~+~K)
~-~a^3V(\Phi)\biggr]
\eqno(3)
$$
where ${\bar{\cal V}}$ is a time-independent volume factor given by
${{{\cal V}(t_f)}\over {a^3(t_f)}}$, with ${\cal V}(t_f)$ being the
three-dimensional volume of the spatial hypersurface at time $t_f$.

Since the field equations are not higher-derivative, the action can be 
modified by adding a boundary term to eliminate factors of ${\ddot a}$.

$$\eqalign{{{I^\prime}\over {\bar{\cal V}}}~&=~{I\over {\bar{\cal V}}}~-
~\int~dt~{d\over{dt}}~
\left[2{{e^{-\Phi}}\over {g_4^2}}{\dot a}({\dot a}^2+3K)~+~6a^2{\dot a}
~\right]
\cr
~&=~\int~dt~\biggl[6a(-{\dot a}^2~+~K)~+~{1\over 2}a^3{\dot\Phi}^2~+~
             2{{e^{-\Phi}}\over {g_4^2}}{\dot \Phi}{\dot a}
                                     ({\dot a}^2+3K)~-~a^3V(\Phi)\biggr]
\cr}
\eqno(4)
$$
The conjugate momenta are then
$$P_a~=~-12 a {\dot a}~+~6{{e^{-\Phi}}\over {g_4^2}}{\dot \Phi}
({\dot a}^2~+~K)
~~~~~~~~~~~                                                                 
P_\Phi~=~a^3{\dot \Phi}~+~2{{e^{-\Phi}}\over {g_4^2}}{\dot a}({\dot a}^2
~+~3K)
\eqno(5)
$$

Expanding in powers of ${{e^{-\Phi}}\over {g_4^2}}$, expressions for ${\dot a}$
and ${\dot \Phi}$ to first order may be obtained
$$\eqalign{{\dot a}\simeq & -{1\over {12a}}P_a+{{e^{-\Phi}}\over {g_4^2}}
{1\over {2a^4}}P_\Phi\left({1\over {144a}}P_a{1\over a}
P_a + K\right)
\cr
{\dot \Phi}\simeq &~{1\over {a^3}}P_\Phi+{{e^{-\Phi}}\over {g_4^2}}
{1\over {6a^4}} P_a \left({1\over {144a}}P_a {1\over a}P_a+ 3K\right)
\cr}
\eqno(6)
$$
and the Hamiltonian is
$$\eqalign{H~\simeq~-6a & \biggl[\left[-{1\over {12a}}P_a~+~
{{e^{-\Phi}}\over {g_4^2}}{1\over {2a^4}}
P_\Phi \left({1\over{12a}} P_a{1\over {12a}} P_a
~+~K\right) \right]^2~+~K \biggr]
\cr
&+6 {{e^{-\Phi}}\over {g_4^2}}~\biggl[{1\over {a^3}} P_\Phi
~+~{{e^{-\Phi}}\over {g_4^2}} {1\over {6a^4}} P_a
\left({1\over {12a}}P_a {1\over {12a}} P_a~+~3K\right)
\biggr]
\cr
&~~~\cdot~\biggl\{ \left[-{1\over {12a}}P_a~+~{{e^{-\Phi}}\over {g_4^2}}
{1\over {2a^4}} P_\Phi \left({1\over {12a}}P_a
{1\over {12a}}P_a~+~K\right)\right]^2~+~K  \biggr\}
\cr
&~~~~~\cdot~\left[-{1\over {12a}} P_a~+~{{e^{-\Phi}}\over {g_4^2}}
{1\over {2a^4}} P_\Phi \left({1\over {12a}} P_a
{1\over {12a}} P_a~+~K\right)\right]
\cr
&~~+~{{a^3}\over 2}~\left[{1\over {a^3}} P_\Phi~+~
{{e^{-\Phi}}\over {g_4^2}}{1\over {6a^4}}P_a
\left({1\over {12a}} P_a{1\over {12a}} 
P_a~+~3K\right)\right]^2~+~a^3 V(\Phi)
\cr}
\eqno(7)
$$

In a Lorentzian space-time, the differential operator representing the
Hamiltonian is obtained by making the substitutions 
$P_a \to -i {\partial\over {\partial a}}$ and
$P_\Phi \to -i {\partial\over {\partial \Phi}}$.  Choosing the 
operator-ordering parameter to be equal to $-1$, so that 
$\left({1\over a}P_a\right)^2 \to -{1\over a}{\partial\over {\partial a}}
{1\over a}{\partial\over {\partial a}}$, the $\hbar \to 0$
limit produces the equation
$$H_0\Psi~=~\left({1\over {24}}{\partial\over {\partial a}}{1\over a}
                       {\partial\over {\partial a}}~-~{1\over {2a^3}}
                            {{\partial^2}\over {{\partial \Phi}^2}}~-~6aK
                          ~+~a^3V(\Phi)~\right) \Psi~=~0
\eqno(8)
$$
This is equivalent to the standard Wheeler-DeWitt equation for standard 
gravity plus a scalar field
$$\eqalign{\left[{1\over {a^p}}{\partial\over {\partial a}} a^p 
{\partial\over {\partial a}}~-~{1\over{a^2}} {\partial^2\over
{\partial \phi^2}}~-~a^2 U(a,\phi)\right] \Psi~&=~0
\cr
U(a,\phi)~=~1~-~a^2~V(\phi)~~~~~~~~~~~V(\phi)~&=~scalar~potential
\cr}
\eqno(9)
$$
where, as above, the operator-ordering parameter $p$ is set equal to $-1$,
using the rescalings 
\hfil\break
$a^2~\to~{1\over {12 {\sqrt K}}}a^2$, 
$\Phi~\to~{1\over {2 {\sqrt 3}}}~\Phi$ and $V(\Phi)~\to~{72~K^{3\over 2}}
V(\Phi)$. 

While this choice of operator ordering is convenient for obtaining 
closed solutions to the Wheeler-DeWitt equation (8), 
other values of this parameter also 
can be used.  The parameter $p$ impacts on the regularity of the wave 
function $\Psi(\Phi,a)$ as $a\to 0$; when $p \ge 1$, regularity in the 
$a\to 0$ limit requires the no-boundary wavefunction [1], whereas the 
divergence in the tunneling wavefunction in the limit $a \to 0$ is 
regulated by a prefactor when $p \le 0$ [34].  Non-singular solutions with 
Lorentzian signature will occur if regularity is maintained  in this limit.

There are additional operator-ordering ambiguities at order 
$O\left({{e^{-\Phi}}\over {g_4^2}}\right)$, because terms containing the 
cube of ${1\over a}P_a$ and the product of ${{e^{-\Phi}}\over {g_4^2}}$ 
and $P_\Phi$ are included in the Hamiltonian.  Suppose 
$\left({1\over a}P_a\right)^2$
is replaced by $-{1\over {a^{p+2}}}{\partial\over {\partial a}}a^p
{\partial\over {\partial a}}=-{1\over a}{1\over {a^{p+1}}}
{\partial\over {\partial a}}{1\over {a^{1-(p+1)}}}
{\partial\over {\partial a}}$  
to obtain the second-order Wheeler-DeWitt 
equation (9), where ${1\over {a^{p+1}}}{\partial\over {\partial a}}
{1\over {a^{1-(p+1)}}}$ represents the ordering of the product of 
${1\over a}$ and ${\partial\over {\partial a}}$.  
The second expression provides 
a way of obtaining higher powers of ${1\over a}P_a$ through iteration,
without the introduction of non-derivative terms.  
For example, $\left({1\over a}P_a\right)^3\to (-i)^3{1\over a}
\left({1\over {a^{p+1}}}
{\partial\over {\partial a}}a^p\right)
\left({1\over {a^{p+1}}}{\partial\over {\partial a}}a^p\right)
{\partial\over {\partial a}}= i{1\over {a^{p+2}}}{\partial\over {\partial a}}
{1\over a}{\partial\over {\partial a}}a^p{\partial\over {\partial a}}$ 
represents a substitution for $\left({1\over a}P_a\right)^3$ which 
does not involve non-derivative terms and therefore has a form similar to the
differential operator in equation (9).

More generally, powers of ${1\over a}$ can be placed to the right of
an operator product containing $P_a$, giving rise to non-derivative
terms.  Indeed, one can consider an arbitrary linear combination of the 
twenty operators ${\cal O}_i$, not all linearly independent, obtained by 
permutation of the factors in the product 
${1\over a}P_a{1\over a}P_a{1\over a}P_a$.  This combination 
$\sum_{i=1}^{20}~\alpha_i {\cal O}_i$ can be shown to be 
equal to $i{1\over {a^p}}{\partial\over {\partial a}}{1\over {a^q}}
{\partial\over {\partial a}}{1\over {a^r}}{\partial\over {\partial a}}
{1\over {a^s}}$, where $p,~q,~r$ and $s$ may be expressed in terms of the
coefficients $\alpha_i$, and $p+q+r+s = 3$.  Similarly, 
${{e^{-\Phi}}\over {g_4^2}}P_\Phi$ may be replaced by 
$-i\left[{{e^{-\Phi}}\over {g_4^2}}{\partial\over {\partial \Phi}}+
t{{e^{-\Phi}}\over {g_4^2}}\right]$, implying the existence of another
operator-ordering parameter $t$ in the Wheeler-DeWitt equation $H\Psi=0$.

The powers $p,~q,~r$ and $s$ are constrained by consistency with
the algebra of supersymmetry constraints and hermiticity.  
While the potential in the heterotic string effective action is also 
determined by supersymmetry, it would be modified by the addition of any 
term not involving the derivative with respect to $a$ in the expansion of
$\left({1\over a}P_a\right)^3$.          

Based on the substitutions
$$\eqalign{{1\over a}P_a~&\to~{{-i}\over {a^{p_1+1}} }
{\partial\over {\partial a}}a^{p_1}
\cr
\left({1\over a}P_a\right)^2~&\to~-{1\over {a^{p_2}} }
{\partial\over {\partial a}}
{1\over {a^{q_2}}}{\partial\over {\partial a}}{1\over {a^{r_2}}}
~~~~~~~~~~~~~~p_2+q_2+r_2~=~2
\cr
\left({1\over a}P_a\right)^3~&\to~{i\over {a^{p_3}}}
{\partial\over {\partial a}}
{1\over {a^{q_3}}}{\partial\over {\partial a}}{1\over {a^{r_3}}}
{\partial\over {\partial a}}{1\over {a^{s_3}}}~~~~~~~p_3+q_3+r_3+s_3=3
\cr}
\eqno(10)
$$
the non-derivative term is

A specific operator-ordering, such as normal ordering, should be used 
uniformly for all of the terms in a Lagrangian field theory.  In this
case, setting ${\hat p}$ equal to $p_1$, it follows that 
$p_2={\hat p}+1,q_2=1,r_2=-{\hat p},p_3={\hat p}+1,q_3=1,r_3=1,
s_3=-{\hat p}$ so that the non-derivative term becomes

$$\eqalign{\bigg\{&{1\over {24}}{{r_2(q_2+r_2+1)}\over {a^3}}
+t{{e^{-\Phi}}\over {g_4^2}} \biggl[ {1\over {864 a^9}}
\biggl[s_3(s_3+r_3+1)(q_3+r_3+s_3+2)+6(q_3+r_3+s_3)
\cr
&\cdot (r_3+s_3(r_3+s_3+2))+{3\over 2}[(r_3+s_3(r_3+s_3+2))(q_3+r_3+s_3+1)+
s_3(r_3+s_3+1)]
\cr
&-6r_2(r_2+q_2+1)+30\biggr]+{K\over 4}{{(7-4p_1)}\over {a^5}}\biggr]
-{{e^{-\Phi}}\over {g_4^2}}{{s_3(r_3+s_3+1)(q_3+r_3+s_3+2)}\over
{1728~a^9}}\biggr\}~\Psi 
\cr}
\eqno(11)
$$
  
A specific operator-ordering, such as normal ordering, should be used
uniformly for all of the terms in a Lagrangian field theory.  In this case, 
setting $p_1$ equal to ${\hat p}$, it follows that $p_2={\hat p}+1,q_2+1,
r_2=-{\hat p},p_3={\hat p}+1,q_3=1,r_3=1,s_3=-{\hat p}$ so that the
non-derivative term becomes

$$\eqalign{\biggl\{{1\over {24}}&{{{\hat p}({\hat p}-2)}\over {a^3}}
+t{{e^{-\Phi}}\over {g_4^2}}\left[{1\over {864~a^9}} \left(
-{{17}\over 2}{\hat p}^3+{{81}\over 2}{\hat p}^2-56{\hat p}+{{93}\over 2}
\right)+{K\over 4}{{(7-4{\hat p})}\over {a^5}}\right]
\cr
&~~~~~~~~~~~~~~~~~~~~~~~~~~~~~~~~~~~~~~~~~~~~~~~~~~~
+{{e^{-\Phi}}\over {g_4^2}} { {{\hat p}({\hat p}-2)({\hat p}-4)}\over
{1728~a^9}}\biggr\}~\Psi
\cr}
\eqno(12)
$$
            
When the factor of $\hbar$ is included in $P_a$, the first term in the 
expression (12) has the same order as $U(a,\Phi)$, whereas the second term 
represents an $O(\hbar)$ contribution.  While this does represent a 
modification of $U(a,\Phi)$, invariance of the entire Lagrangian under 
supersymmetry transformations implies that consistency with the 
supersymmetry constraints would not be affected by rearrangement of terms 
in the Hamiltonian.   The $O(1)$ part of the non-derivative term should be
set equal to zero to minimize the correction to $U(a,\Phi)$,
and this requires ${\hat p}=0,2$, when the operator-ordering (10)
is applied uniformly.  In terms of the original operator-ordering parameter
$p$, these values correspond to $p=-1,2$.  Non-derivative terms can
be removed at $O(1)$ for more general values of $p$ only if the
substitution $\left({1\over a}P_a\right)^2 \to -{1\over {a^{p+2}}}
{\partial\over {\partial a}}a^p{\partial\over {\partial a}}$ is used
instead of  $\left({1\over a}P_a\right)^2 \to -{1\over {a^{{\hat p}+1}}}
{\partial\over {\partial a}}{1\over a}{\partial\over {\partial a}}a^{\hat p}$.
  
The heterotic string potential is
$$\eqalign{V~&=~e^{-{\cal K}}~[{\cal K^\alpha}
\left({\cal K}^{-1}\right)_\alpha^\beta {\cal K}_\beta~+~3]
\cr
{\cal K}~&=~ln(S~+~{\bar S})~+~3~ln~(T~+~{\bar T})~-~ln~\vert {\cal W}(S)
\vert^2
~~~~~~~~~~~~{\cal K}_\alpha~=~{{\partial {\cal K}}\over 
{\partial \phi^\alpha}}
\cr
Re~S~&=~{{e^{-\Phi}}\over {g_4^2}}
~~~~~~~~~~~~Re~T~=~e^\sigma
\cr}
\eqno(13)
$$
where $\sigma$ is set equal to a constant, $\sigma_0$, and 
$\{\phi^\alpha\}$ represents the fields $S$ and $T$, with the
other matter fields set equal to zero [35] - [38], and ${\cal W}(S)$ is the 
superpotential.

The equation $H\Psi=0$ can be solved approximately by using 
$\epsilon \equiv {{e^{-\Phi}}\over {g_4^2}}$ as an expansion parameter 
and noting that $\Psi=\Psi_0+\epsilon\Psi_1$ is a solution to
$(H_0+\epsilon H_1)\Psi=0$ to order $O(\epsilon)$ if 
$H_0\Psi_1=-H_1\Psi_0$.  Given the Hamiltonian (7), the first-order 
correction is 

$$\eqalign{-{1\over 4}{\partial\over {\partial a}}{1\over {a^4}}&
{\partial\over {\partial \Phi}} \left(-{1\over {144a}}{\partial
\over {\partial a}} {1\over a} {\partial\over {\partial a}}~+~K\right)
-~{1\over {4a^3}}{\partial\over {\partial\Phi}}\left(-{1\over {144a}}
{\partial\over {\partial a}} {1\over a} {\partial\over {\partial a}}~+~K
\right){1\over a} {\partial\over {\partial a}}
\cr
+{1\over {2a^3}} {\partial\over {\partial \Phi}}&\left(-{1\over {144a}}
{\partial\over {\partial a}} {1\over a} {\partial\over {\partial a}}
+K\right){1\over a}{\partial\over {\partial a}}
-{1\over {12}}\left({\partial\over {\partial\Phi}}-1\right)
{1\over {a^4}} {\partial\over {\partial a}}
\left(-{1\over {144a}} {\partial\over {\partial a}} {1\over a}
{\partial\over {\partial a}}~+~3K\right)
\cr
&~~~~-~{1\over {12a}}{\partial\over {\partial a}}
\left(-{1\over {144a}}{\partial\over {\partial a}} 
{1\over a} {\partial\over {\partial a}}~+~3K\right)
{1\over {a^3}} {\partial\over {\partial\Phi}}
\cr}
\eqno(14)
$$
and

$$\eqalign{H_1 \Psi_0~&=~{1\over {a^4}}\left({K\over 4}~-~{1\over {576~a^4}}
\right)
{{\partial\Psi_0}\over {\partial a}}~+~{1\over {576~a^7}}{{\partial^2\Psi_0}
\over {\partial a^2}}~-~{1\over {1728~a^6}}{{\partial^3\Psi_0}\over
{\partial a^3}}
\cr
&~~~~~~+~{1\over {a^5}}\left({{7K}\over 4}-{{35}\over {576~a^4}}\right)
{{\partial \Psi_0}\over {\partial \Phi}}
~+~{1\over {24~a^8}}
{{\partial^2\Psi_0}\over {\partial a\partial\Phi}}
\cr
&~~~~~~~~-~{1\over {64~a^7}}
{{\partial^3\Psi_0}\over {\partial a^2 \partial\Phi}}
~+~{1\over {864~a^6}}{{\partial^4\Psi_0}\over {\partial a^3\partial\Phi}}
\cr}
\eqno(15)
$$

In the classically allowed range, the `no boundary' wavefunction [1], for 
example, is
$$\Psi_{0_{NB}}~\simeq~exp\left({{24K^{3\over 2}}\over V}\right)~
\left({{a^2V}\over {6K}}~-~1\right)^{-{1\over 4}} cos\biggl[{{24K^{3\over 2}}
\over V}\left({{a^2V}\over {6K}}~-~1\right)^{3\over 2}~-~{\pi\over 4}
\biggr] 
\eqno(16)
$$
and

$$\eqalign{H_1 \Psi_{0_{NB}}=
&{1\over {a^3}} exp\left({{24K^{3\over 2}}\over V}\right)
\biggl[{5\over {24}}\left({V\over {24K}}\right)^3 \left[{{a^2V}\over {6K}}-1
\right]^{-{{13}\over 4}}+\left({{a^2V}\over {6K}}-2\right){V\over {48}}
\cr
&~~~~~~~~~~~~~~~~~~\cdot\left[{{a^2V}\over {6K}}-1\right]^{-{5\over 4}}\biggr]
\cdot~cos~\left[{{24K^{3\over 2}}\over V}\left({{a^2V}\over {6K}}
~-~1\right)^{3\over 2}~-~{\pi\over 4}\right]
\cr
&+{1\over {a^3}}
 exp \left({{24K^{3\over 2}}\over V}\right)
\biggl[{5\over {K^{3\over 2}}}\left({V\over {144}}\right)^2
\left[{{a^2V}\over {6K}}-1\right]^{-{7\over 4}}
-\left({{a^2V}\over {6K}}+~2\right)K^{3\over 2}
\cr
&~~~~~~~~~~~~~~~~~~~~~\cdot\left[{{a^2V}\over {6K}}-1\right]^{1\over 4}\biggr]
~\cdot~sin \left[{{24K^{3\over 2}}\over V}\left({{a^2V}\over {6K}}
~-~1\right)^{3\over 2}~-~{\pi\over 4}\right]
\cr
+\Bigg\{{{V^\prime}\over {a^5}}&
\left({{7K}\over 4}-{{35}\over {576a^4}}\right)~
~\left[-{{a^2}\over {24K}}\left[{{a^2V}\over {6K}}-1\right]^{-{5\over 4}} 
-{{24K^{3\over 2}}\over {V^2}}\left[{{a^2V}\over {6K}}-1\right]^{-{1\over 4}}
\right]
\cr
+{{V^\prime}\over {24a^8}}&
\Bigg[{5\over {288}}{{a^3V}\over {K^2}}\left[{{a^2V}\over {6K}}-1
\right]^{-{9\over 4}}+\left(2{{aK^{1\over 2}}\over V}-{a\over {12K}}\right)
\left[{{a^2V}\over {6K}}-1\right]^{-{5\over 4}}
\cr
&~~~~~~~~~~~~~~~~~~~~~~~~~~~~~~~~~~-72{{a^3K}\over V}
\left[{{a^2V}\over {6K}}-1\right]^{3\over 4}+288{{aK^2}\over {V^2}} 
\left[{{a^2V}\over {6K}}-1\right]^{7\over 4}\Bigg]
\cr}
$$
$$\eqalign{
&-{{V^\prime}\over {64a^7}}
\Bigg[-{5\over {384}}{{a^4V^2}\over {K^3}}\left[{{a^2V}\over 
{6K}}-1\right]^{-{{13}\over 4}}+\left({{25}\over {288}}
{{a^2V}\over {K^2}}-{5\over 6}{{a^2}\over {K^{1\over 2}}}
\right)\left[{{a^2V}\over {6K}}-1\right]^{-{9\over 4}}
\cr
&~~~+\left(2{{K^{1\over 2}}\over V}-{1\over {12K}}\right)
\left[{{a^2V}\over {6K}}-1\right]^{-{5\over 4}}-18a^4\left[{{a^2V}\over 
{6K}}-1\right]^{-{1\over 4}}
+\left(3456{{a^2K^{5\over 2}}\over {V^2}}-72{{a^2K}\over V}\right)
\cr
&~~~~\cdot\left[{{a^2V}\over {6K}}-1\right]^{3\over 4}+
{{288K^2}\over {V^2}}\left[{{a^2V}\over {6K}}-1\right]^{7\over 4}\Bigg]
+{{V^\prime}\over {864 a^6}}\biggl[{{65}\over {4608}}{{a^5V^3}\over {K^4}}
\left[{{a^2V}\over {6K}}-1\right]^{-{{17}\over 4}}
\cr
&~~~+\left({5\over 8}{{a^3V}\over {K^{3\over 2}}}-{{15}\over {128}}
{{a^3V^2}\over {K^3}}\right)
\left[{{a^2V}\over {6K}}-1\right]^{-{{13}\over 4}}
+\left({5\over {24}}{{aV}\over {K^2}}-{5\over 2}{a\over {K^{1\over 2}}}\right) 
\left[{{a^2V}\over {6K}} -1\right]^{-{9\over 4}}
\cr
&-{{a^5V}\over K}\left[{{a^2V}\over {6K}}-1\right]^{-{5\over 4}}
+\left(864{{a^3K^{5\over 2}}\over V}-80a^3 \right)
\left[{{a^2V}\over {6K}}-1\right]^{-{1\over 4}}
\cr
&
+10368{{a^2K^{5\over 2}}\over {V^2}}
\left[{{a^2V}\over {6K}}-1\right]^{3\over 4}
+10368{{a^5K^2}\over V}\left[{{a^2V}\over {6K}}-1\right]^{7\over 4}
\cr
&-41472{{a^3K^3}\over {V^2}}\left[{{a^2V}\over {6K}}-1\right]^{{11}\over 4}
\Bigg]\Bigg\}
\cdot~exp\left({{24K^{3\over 2}}\over V}\right)~cos~
\left[{{24K^{3\over 2}}\over V}\left({{a^2V}\over {6K}}-1
\right)^{3\over 2}-{\pi\over 4}\right]
\cr
&+\bigg\{ {{V^\prime}\over {a^5}}
\left({{7K}\over 4}-{{35}\over {576a^4}}\right)
~\biggl[-{{6a^2K^{1\over 2}}\over V}~\left[{{a^2V}\over {6K}}-1
\right]^{1\over 4}+{{24K^{3\over 2}}\over {V^2}}~
\left[{{a^2V}\over {6K}}-1\right]^{5\over 4}\biggr]
\cr
&+{{V^\prime}\over {24a^8}}\left(288{{aK^2}\over {V^2}}
-2{{aK^{1\over 2}}\over V}\right)\left[{{a^2V}\over {6K}}-1\right]^{1\over 4}
-{{V^\prime}\over {64a^7}}
\biggl[-{5\over {24}}{{a^4V}\over {K^{3\over 2}}}\left[{{a^2V}\over {6K}}
-1\right]^{-{7\over 4}}
\cr
&+{5\over 6}{{a^2}\over {K^{1\over 2}}}\left[{{a^2V}\over {6K}}-1
\right]^{-{3\over 4}}
+\left(288{{K^2}\over {V^2}}-2{{K^{1\over 2}}\over V}\right)
\left[{{a^2V}\over {6K}}-1\right]^{1\over 4}+864{{a^4K^{3\over 2}}\over V}
\left[{{a^2V}\over {6K}}-1\right]^{5\over 4}
\cr
&-3456{{a^2K^{5\over 2}}\over {V^2}}\left[{{a^2V}\over {6K}}
-1\right]^{9\over 4}\biggr]
+{{V^\prime}\over {864a^6}}\biggl[{5\over {18}}{{a^5V^2}\over 
{K^{5\over 2}}}\left[{{a^2V}\over {6K}}-1\right]^{-{{11}\over 4}}
\cr
&+\left(10a^3-{{25}\over {12}}{{a^3V}\over {K^{3\over 2}}}\right)
\left[{{a^2V}\over {6K}}-1\right]^{-{7\over 4}}+{5\over 2}
{a\over {K^{1\over 2}}}\left[{{a^2V}\over {6K}}-1\right]^{-{3\over 4}}
\cr
&+576a^5K^{1\over 2}\left[{{a^2V}\over {6K}}-1\right]^{1\over 4}
+\left(1728{{a^3K^{3\over 2}}\over V}-41472{{a^3K^3}\over {V^2}}\right)
\left[{{a^2V}\over {6K}}-1\right]^{5\over 4}
\cr
&-10368{{aK^{5\over 2}}\over {V^2}}\left[{{a^2V}\over {6K}}-1\right]^{9\over 4}
\biggr]\bigg\}\cdot~exp\left({{24K^{3\over 2}}\over V}\right)
~sin\left[{{24K^{3\over 2}}\over V}\left({{a^2V}\over {6K}}
-1\right)^{3\over 2}-{\pi\over 4}\right]
\cr}
\eqno(17)
$$
where terms involving the derivative of the potential with respect to 
$\Phi$ may be evaluated using the operator (15).

In the classically forbidden region [1][2], 
$$\Psi_{0_{NB}}~\simeq~{1\over 2}~
\left[1~-~{{a^2V}\over {6K}}\right]^{-{1\over 4}}
~\cdot~ exp\left[{{24K^{3\over 2}}
\over V}~\left(1~-~\left(1~-~{{a^2V}\over {6K}}\right)^{3\over 2}\right)
\right]
\eqno(18)
$$
and

$$\eqalign{H_1 \Psi_{0_{NB}}~&\simeq~{1\over 2}~
\biggl\{-{5\over {24}}\left({V\over {24K}}\right)^3
~\left[1~-~{{a^2V}\over {6K}}\right]^{-{{13}\over 4}}
~-~{5\over {K^{3\over 2}}}\left({V\over {144}}\right)^2~
\left[1~-~{{a^2V}\over {6K}}\right]^{-{7\over 4}}
\cr
&~~~~~~~~~~~~~~+~{V\over {48}}~\left(2-{{a^2V}\over {6K}}\right)
\left[1-{{a^2V}\over {6K}}\right]^{-{5\over 4}}
+K^{3\over 2}\left({{a^2V}\over {6K}}+2\right)
\left[1-{{a^2V}\over {6K}}\right]^{1\over 4}\biggr\}
\cr
&~~~~~~~~~~~~~\cdot~exp\left({{24K^{3\over 2}}\over V}
\left[1~-~\left(1~-~{{a^2V}\over {6K}}\right)^{3\over 2}\right)\right]
\cr
&+\Bigg\{{{V^\prime}\over {2a^5}}\left({{7K}\over 4}
-{{35}\over {576a^4}}\right)\bigg[{{a^2}\over {24K}}
\left[1-{{a^2V}\over {6K}}\right]^{-{5\over 4}}-{{24K^{3\over 2}}\over {V^2}}
\left[1-{{a^2V}\over {6K}}\right]^{-{1\over 4}}
\cr
&~~~~~~~~~~~~~~~~~~~~~~~~+6aK^{1\over 2}\left[1-{{a^2V}\over {6K}}
\right]^{1\over 4}+{{24K^{3\over 2}}\over {V^2}}\left[1-{{a^2V}
\over {6K}}\right]^{5\over 4}\bigg]
\cr
&+{{V^\prime}\over {48a^8}}\bigg[{5\over {288}}{{a^3V}\over {K^2}}
\left[1-{{a^2V}\over {6K}}\right]^{-{9\over 4}}+\left({a\over {12K}}-
2{{aK^{1\over 2}}\over V}\right)\left[1-{{a^2V}\over {6K}}
\right]^{-{5\over 4}}
\cr
&+\left(2{{aK^{1\over 2}}\over {V^2}}-288{{aK^2}\over {V^2}}\right)
\left[1-{{a^2V}\over {6K}}\right]^{1\over 4}+72{{a^3K^2}\over {V^2}}
\left[1-{{a^2V}\over {6K}}\right]^{3\over 4}
\cr
&~~~+288{{aK^2}\over {V^2}}\left[1-{{a^2V}\over {6K}}\right]^{7\over 4}\bigg]
-{{V^\prime}\over {128a^7}}\bigg[{5\over {384}}{{a^4V^2}\over {K^3}}
\left[1-{{a^2V}\over {6K}}\right]^{-{{13}\over 4}}
\cr
&~~~+\left({{25}\over {288}}{{a^2V}\over {K^2}}-{5\over 6}
{{a^2}\over {K^{1\over 2}}}\right)
\left[1-{{a^2V}\over {6K}}\right]^{-{9\over 4}}
+{5\over {24}}{{a^4V}\over {K^{3\over 2}}}
\left[1-{{a^2V}\over {6K}}\right]^{-{7\over 4}}
\cr
+&\left({1\over {12K}}-2{{K^{1\over 2}}\over V}\right)
\left[1-{{a^2V}\over {6K}}\right]^{-{5\over 4}}
+{5\over 6}{{a^2}\over {K^{1\over 2}}}
\left[1-{{a^2V}\over {6K}}\right]^{-{3\over 4}}
-18a^4\left[1-{{a^2V}\over {6K}}\right]^{-{1\over 4}}
\cr
&~~~+\left(2{{K^{1\over 2}}\over V}-288{{K^2}\over {V^2}}\right)
\left[1-{{a^2V}\over {6K}}\right]^{1\over 4}+\left(72{{a^2K}\over V}
-3456{{a^2K^{5\over 2}}\over {V^2}}\right)\left[1-{{a^2V}\over {6K}}
\right]^{3\over 4}
\cr
&~~~+864{{a^4K^{3\over 2}}\over V}\left[1-{{a^2V}\over {6K}}\right]^{5\over 4}
+288{{K^2}\over {V^2}}\left[1-{{a^2V}\over {6K}}\right]^{7\over 4}
+3456{{a^2K^{5\over 2}}\over {V^2}}\left[1-{{a^2V}\over {6K}}
\right]^{9\over 4}\bigg] 
\cr
&+{{V^\prime}\over {1728a^8}}\bigg[{{65}\over {4608}}
{{a^5V^3}\over {K^4}}\left[1-{{a^2V}\over {6K}}\right]^{-{{17}\over 4}}+
\left({{15}\over {128}}{{a^3V^2}\over {K^3}}-{5\over 8}{{a^3V}\over 
{K^{3\over 2}}}\right)\left[1-{{a^2V}\over {6K}}\right]^{-{{13}\over 4}}
\cr}
$$
$$\eqalign{
&~~~+{5\over {18}}{{a^5V^2}\over {K^{5\over 2}}}
\left[1-{{a^2V}\over {6K}}\right]^{-{{11}\over 4}}
+\left({5\over {24}}{{aV}\over {K^2}}-{5\over 2}{a\over {K^{1\over 2}}}
\right)\left[1-{{a^2V}\over {6K}}\right]^{-{9\over 4}}
\cr
&~~~+\left({{25}\over {12}}{{a^3V}\over {K^{3\over 2}}}-10a^3\right)
\left[1-{{a^2V}\over {6K}}\right]^{-{7\over 4}}
-{{a^5V}\over K}\left[1-{{a^2V}\over {6K}}\right]^{-{5\over 4}}
\cr
&~~~+{5\over 2}{a\over {K^{1\over 2}}}
\left[1-{{a^2V}\over {6K}}\right]^{-{3\over 4}}+
\left(864{{a^3K^{3\over 2}}\over V}-80a^3\right)
\left[1-{{a^2V}\over {6K}}\right]^{-{1\over 4}}
\cr
&~~~-576a^5K^{1\over 2}\left[1-{{a^2V}\over {6K}}\right]^{1\over 4}
-10368{{aK^{5\over 2}}\over {V^2}}\left[1-{{a^2V}\over {6K}}\right]^{3\over 4}
\cr
&~~~+\left(1728{{a^3K^{3\over 2}}\over V}-41472{{a^3K^3}\over {V^2}}\right)
\left[1-{{a^2V}\over {6K}}\right]^{5\over 4}
+10368{{a^5K^2}\over V} \left[1-{{a^2V}\over {6K}}\right]^{7\over 4}
\cr
&~~~+10368{{aK^{5\over 2}}\over {V^2}} 
\left[1-{{a^2V}\over {6K}}\right]^{9\over 4}
+41472{{a^3K^3}\over {V^2}}\left[1-{{a^2V}\over {6K}}\right]^{{11}\over 4}
\bigg]\Bigg\}
\cr
&~~~~~~~~~~~~~~~~~~~~~~~~~~~~~~~~~~~~~~~~~
\cdot~exp\left[{{24K^{3\over 2}}\over V}\left(1-\left(1-{{a^2V}\over {6K}}
\right)^{3\over 2}\right)\right]
\cr}
\eqno(19)
$$

Similarly, the tunneling wave function in the classically acceptable 
region [39]-[41] is
$$\Psi_{0_T}~\simeq~2~e^{{i\pi}\over 4}
\left[{{a^2 V}\over {6K}}~-~1\right]^{-{1\over 4}} 
exp\left[{{24K^{3\over 2}}\over V}\left(1-
i\left({{a^2V}\over {6K}}~-~1\right)^{3\over 2}\right)\right]
\eqno(20)
$$
and

$$\eqalign{&H_1\Psi_{0_T}\simeq{2\over {a^3}} e^{{i\pi}\over 4} 
\biggl\{{5\over {24}}\left({V\over {24K}}\right)^3
\left[{{a^2V}\over {6 K}}-1\right]^{-{{13}\over 4}}
+{{5i}\over {K^{3\over 2}} }\left({V\over {144}}\right)^2   
\left[{{a^2V}\over {6K}}-1\right]^{-{7\over 4}}
\cr
&~~~~~~~~~~~~~~+{V\over {48}}\left(2-{{a^2V}\over {6K}}\right)
\left[{{a^2V}\over {6K}}-1\right]^{-{5\over 4}}
-iK^{3\over 2}\left({{a^2V}\over {6K}}+2\right)
\left[{{a^2V}\over {6K}}-1\right]^{1\over 4}\biggr\}
\cr
&~~~~~~~~~~~~~~~~~~~~~~~~~~~~~~~~~~~~~~~\cdot~exp\left[{{24K^{3\over 2}}
\over V}\left(1-i\left({{a^2V}\over {6K}}-1\right)^{3\over 2}\right)\right]
\cr
&+\Bigg\{{{2V^\prime}\over {a^5}}\left({{7K}\over 4}-{{35}\over 
{576a^4}}\right)e^{{i\pi}\over 4}
\Bigg[-{{a^2}\over {24K}}\left[{{a^2V}\over {6K}}-1
\right]^{-{5\over 4}}-{{24K^{3\over 2}}\over {V^2}}\left[{{a^2V}\over {6K}}
-1\right]^{-{1\over 4}}
\cr
&~~~-6i{{a^2K^{1\over 2}}\over V}\left[{{a^2V}\over
{6K}}-1\right]^{1\over 4}+24iK^{3\over 2}\left[{{a^2V}\over {6K}}-1
\right]^{5\over 4}\biggr]+{{V^\prime}\over {12a^8}}e^{i{\pi\over 4}}
\Bigg[{5\over {288}}{{a^3V}\over {K^2}}\left[{{a^2V}\over {6K}}-1
\right]^{-{9\over 4}}
\cr
&~~~+\left(2{{aK^{1\over 2}}\over V}-{a\over {12K}}\right)\left[{{a^2V}\over
{6K}}-1\right]^{-{5\over 4}}
+\left(288i{{aK^2}\over {V^2}}-2i{{aK^{1\over 2}}
\over V}\right)\left[{{a^2V}\over {6K}}-1\right]^{1\over 4}
\cr}
$$
$$\eqalign{
&~-72{{a^3K}\over V}\left[{{a^2K}\over {6K}}-1\right]^{3\over 4}+
288{{aK^2}\over {V^2}}\left[{{aK^2}\over V}-1\right]^{7\over 4}\Bigg]
-{{V^\prime}\over {32a^7}}e^{i{\pi\over 4}}\Bigg[-{5\over {384}}
{{a^4V^2}\over {K^3}}
\cr
&~~~\cdot\left[{{a^2V}\over {6K}}-1\right]^{-{{13}\over 4}}
+\left({{25}\over {288}}{{a^2V}\over {K^2}}-{5\over 6}{{a^2}\over 
{K^{1\over 2}}}\right)\left[{{a^2V}\over {6K}}-1\right]^{-{9\over 4}}
-{{5i}\over {24}}{{a^4V}\over {K^{3\over 2}}} \left[{{a^2V}\over {6K}}-1
\right]^{-{7\over 4}}
\cr
&~~~+\left(2{{K^{1\over 2}}\over V}-{1\over {12K}}\right)
\left[{{a^2V}\over {6K}}
-1\right]^{-{5\over 4}}+{{5i}\over 6}{{a^2}\over {K^{1\over 2}}}
\left[{{a^2V}\over {6K}}-1\right]^{-{3\over 4}}
-18a^4\left[{{a^2V}\over {6K}}-1\right]^{-{1\over 4}}
\cr
&~~~+\left(288i{{K^2}\over {V^2}}-2i{{K^{1\over 2}}\over V}\right)
\left[{{a^2V}\over {6K}}-1\right]^{1\over 4}
+\left(3456{{a^2K^{5\over 2}}\over {V^2}}-72{{a^2K}\over V}\right)
\cdot\left[{{a^2V}\over {6K}}-1\right]^{3\over 4}
\cr
&~~~+864i{{a^4K^{3\over 2}}\over V}\left[{{a^2V}\over {6K}}-1\right]^{5\over 4}
+288{{K^2}\over {V^2}}\left[{{a^2V}\over {6K}}-1\right]^{7\over 4}
-3456i{{a^2K^{5\over 2}}\over {V^2}}\left[{{a^2V}\over 
{6K}}-1\right]^{9\over 4}\biggr]
\cr
&+{{V^\prime}\over {432a^6}}\Bigg[{{65}\over {4608}}{{a^5V^3}\over {K^4}}
\left[{{a^2V}\over {6K}}-1\right]^{-{{17}\over 4}}
+\left({5\over 8}{{a^3V}\over {K^{3\over 2}}}-{{15}\over {128}}
{{a^3V^2}\over {K^3}}\right)\left[{{a^2V}\over {6K}}-1\right]^{-{{13}\over 4}}
\cr
&~~~+{{5i}\over {18}}{{a^5V^2}\over {K^{5\over 2}}}\left[{{a^2V}\over 
{6K}}-1\right]^{-{{11}\over 4}}
+\left({5\over {24}}{{aV}\over {K^2}}
-{5\over 2}{a\over {K^{1\over 2}}}\right)\left[{{a^2V}\over {6K}}-1
\right]^{-{9\over 4}}
\cr
&~~~+\left(10ia^3-{{25i}\over {12}}{{a^3V}\over {K^{3\over 2}}}\right)
\left[{{a^2V}\over {6K}}-1\right]^{-{7\over 4}}
-{{a^5V}\over K}\left[{{a^2V}\over {6K}}-1\right]^{-{5\over 4}}
+{{5i}\over 2}{a\over {K^{1\over 2}}}\left[{{a^2V}\over {6K}}-1
\right]^{-{3\over 4}}
\cr
&~~~+\left(864{{a^3K^{3\over 2}}\over V}-80a^3\right)
\left[{{a^2V}\over {6K}}-1\right]^{-{1\over 4}}
+576ia^5K^{1\over 2}\left[{{a^2V}\over {6K}}-1\right]^{1\over 4}
\cr
&~~~+10368{{aK^{5\over 2}}\over V}
\cdot\left[{{a^2V}\over {6K}}-1\right]^{3\over 4}
+\left(1728i{{a^3K^{3\over 2}}\over V}-41472i{{a^3K^3}\over {V^2}}\right)
\left[{{a^2V}\over {6K}}-1\right]^{5\over 4}
\cr
&~~~+10368{{a^5K^2}\over V}\left[{{a^2V}\over {6K}}-1\right]^{7\over 4}
-10368i{{aK^{5\over 2}}\over V}
\left[{{a^2V}\over {6K}}-1\right]^{9\over 4}
\cr
&~~~~~~~~~~~~~~~~~~~~~~~~
-41472{{a^3K^3}\over {V^2}}\left[{{a^2V}\over {6K}}-1\right]^{{11}\over 4}
\Bigg]\Bigg\}\cdot exp\left[{{24K^{3\over 2}}\over V}
\left(1-i\left({{a^2V}\over {6K}}-1\right)^{3\over 2}\right)\right]
\cr}
\eqno(21)
$$
whereas the tunneling wave function in the classically forbidden region [41]
is

$$\eqalign{\Psi_{0T}~&\simeq~\left[1-{{a^2V}\over {6K}}\right]^{-{1\over 4}}
\biggl\{exp\left[{{24K^{3\over 2}}\over V}\left(1-\left(1-{{a^2V}
\over {6K}}\right)^{3\over 2}\right)\right]
\cr
&~~~~~~~~~~~~~~~~~~~~~~~~~~~~~~~~~~~~~~~~~~+~2i~exp\left[{
{24K^{3\over 2}}\over V}~exp\left(1+\left(1-{{a^2V}\over 
{6K}}\right)^{3\over 2}\right)\right]\biggr\}
\cr}
\eqno(22)
$$
and

$$\eqalign{H_1\Psi_{0T}&\simeq{1\over {a^3}}
\biggl\{-{5\over {24}}\left({{aV}\over
{24K}}\right)^3\left[1-{{a^2V}\over {6K}}\right]^{-{{13}\over 4}}-
{5\over {K^{3\over 2}}}\left({V\over {144}}\right)^2\left[1-{{a^2V}\over {6K}}
\right]^{-{7\over 4}}
\cr
&~~~~+{V\over {48}}\left(2-{{a^2V}\over {6K}}\right)\cdot\left[1-{{a^2V}\over 
{6K}}\right]^{-{5\over 4}}+K^{3\over 2}\left({{a^2V}\over {6K}}+2\right)
\left[1-{{a^2V}\over {6K}}\right]^{1\over 4}\biggr\}
\cr
&~~~~~~~~~~~~~~~~~~~~~~~~~~~~~~~~~~~~~~\cdot~exp\left[{{24K^{3\over 2}}\over V}
\left(1-\left(1-{{a^2V}\over {6K}}\right)^{3\over 2}\right)\right]
\cr
&+~{{2i}\over {a^3}}\Bigg\{-{5\over {24}}\left({{aV}\over {24K}}\right)^3
\left[1-{{a^2V}\over {6K}}\right]^{-{{13}\over 4}}+{5\over {K^{3\over 2}}}
\left({V\over {144}}\right)^2 \left[1-{{a^2V}\over {6K}}\right]^{-{7\over 4}}
\cr
&~~~~{V\over {48}}\left(2-{{a^2V}\over {6K}}\right)\left[1-{{a^2V}\over {6K}}
\right]^{-{5\over 4}}-K^{3\over 2}\left({{a^2V}\over {6K}}+2\right)\left[1-
{{a^2V}\over {6K}}\right]^{1\over 4}\Bigg\}
\cr
&~~~~~~~~~~~~~~~~~~~~~~~~~~~~~~~~~~~~~~\cdot~~exp\left[{{24K^{3\over 2}}\over 
V}\left(1+\left(1-{{a^2V}\over {6K}}\right)^{3\over 2}\right)\right]
\cr
&+\Bigg\{{{V^\prime}\over {a^5}}\left({{7K}\over 4}
-{{35}\over {576a^4}}\right)\bigg[{{a^2}\over {24K}}
\left[1-{{a^2V}\over {6K}}\right]^{-{5\over 4}}-{{24K^{3\over 2}}\over {V^2}}
\left[1-{{a^2V}\over {6K}}\right]^{-{1\over 4}}
\cr
&~~~~~~~~~~~~~~~~~~~~~~~~+6aK^{1\over 2}\left[1-{{a^2V}\over {6K}}
\right]^{1\over 4}+{{24K^{3\over 2}}\over {V^2}}\left[1-{{a^2V}
\over {6K}}\right]^{5\over 4}\bigg]
\cr
&+{{V^\prime}\over {24a^8}}\bigg[{5\over {288}}{{a^3V}\over {K^2}}
\left[1-{{a^2V}\over {6K}}\right]^{-{9\over 4}}+\left({a\over {12K}}-
{{2aK^{1\over 2}}\over V}\right)\left[1-{{a^2V}\over {6K}}
\right]^{-{5\over 4}}
\cr
&+\left(2{{aK^{1\over 2}}\over {V^2}}-288{{aK^2}\over {V^2}}\right)
\left[1-{{a^2V}\over {6K}}\right]^{1\over 4}+72{{a^3K^2}\over {V^2}}
\left[1-{{a^2V}\over {6K}}\right]^{3\over 4}
\cr
&~~~+288{{aK^2}\over {V^2}}\left[1-{{a^2V}\over {6K}}\right]^{7\over 4}\bigg]
-{{V^\prime}\over {64a^7}}\bigg[{5\over {384}}{{a^4V^2}\over {K^3}}
\left[1-{{a^2V}\over {6K}}\right]^{-{{13}\over 4}}
\cr
&~~~+\left({{25}\over {288}}{{a^2V}\over {K^2}}-{5\over 6}
{{a^2}\over {K^{1\over 2}}}\right)
\left[1-{{a^2V}\over {6K}}\right]^{-{9\over 4}}
+{5\over {24}}{{a^4V}\over {K^{3\over 2}}}
\left[1-{{a^2V}\over {6K}}\right]^{-{7\over 4}}
\cr
&~~~+\left({1\over {12K}}-2{{K^{1\over 2}}\over V}\right)
\left[1-{{a^2V}\over {6K}}\right]^{-{5\over 4}}
+{5\over 6}{{a^2}\over {K^{1\over 2}}}
\left[1-{{a^2V}\over {6K}}\right]^{-{3\over 4}}
-18a^4\left[1-{{a^2V}\over {6K}}\right]^{-{1\over 4}}
\cr
&~~~+\left(2{{K^{1\over 2}}\over V}-288{{K^2}\over {V^2}}\right)
\left[1-{{a^2V}\over {6K}}\right]^{1\over 4}+\left(72{{a^2K}\over V}
-3456{{a^2K^{5\over 2}}\over {V^2}}\right)\left[1-{{a^2V}\over {6K}}
\right]^{3\over 4}
\cr
&~~~+864{{a^4K^{3\over 2}}\over V}\left[1-{{a^2V}\over {6K}}\right]^{5\over 4}
+288{{K^2}\over {V^2}}\left[1-{{a^2V}\over {6K}}\right]^{7\over 4}
+3456{{a^2K^{5\over 2}}\over {V^2}}\left[1-{{a^2V}\over {6K}}
\right]^{9\over 4}\bigg] 
\cr}
$$
$$\eqalign{&+{{V^\prime}\over {864a^8}}\bigg[{{65}\over {4608}}
{{a^5V^3}\over {K^4}}\left[1-{{a^2V}\over {6K}}\right]^{-{{17}\over 4}}+
\left({{15}\over {128}}{{a^3V^2}\over {K^3}}-{5\over 8}{{a^3V}\over 
{K^{3\over 2}}}\right)\left[1-{{a^2V}\over {6K}}\right]^{-{{13}\over 4}}
\cr
&~~~+{5\over {18}}{{a^5V^2}\over {K^{5\over 2}}}
\left[1-{{a^2V}\over {6K}}\right]^{-{{11}\over 4}}
+\left({5\over {24}}{{aV}\over {K^2}}-{5\over 2}{a\over {K^{1\over 2}}}
\right)\left[1-{{a^2V}\over {6K}}\right]^{-{9\over 4}}
\cr
&~~~+\left({{25}\over {12}}{{a^3V}\over {K^{3\over 2}}}-10a^3\right)
\left[1-{{a^2V}\over {6K}}\right]^{-{7\over 4}}
-{{a^5V}\over K}\left[1-{{a^2V}\over {6K}}\right]^{-{5\over 4}}
\cr
&~~~+{5\over 2}{a\over {K^{1\over 2}}}
\left[1-{{a^2V}\over {6K}}\right]^{-{3\over 4}}+
\left(864{{a^3K^{3\over 2}}\over V}-80a^3\right)
\left[1-{{a^2V}\over {6K}}\right]^{-{1\over 4}}
\cr
&~~~-576a^5K^{1\over 2}\left[1-{{a^2V}\over {6K}}\right]^{1\over 4}
-10368{{aK^{5\over 2}}\over {V^2}}
\left[1-{{a^2V}\over {6K}}\right]^{3\over 4}
\cr
&~~~+\left(1728{{a^3K^{3\over 2}}\over V}-41472{{a^3K^3}\over {V^2}}\right)
\left[1-{{a^2V}\over {6K}}\right]^{5\over 4}
+10368{{a^5K^2}\over V} \left[1-{{a^2V}\over {6K}}\right]^{7\over 4}
\cr
&~~~+10368{{aK^{5\over 2}}\over {V^2}} 
\left[1-{{a^2V}\over {6K}}\right]^{9\over 4}
+41472{{a^3K^3}\over {V^2}}\left[1-{{a^2V}\over {6K}}\right]^{{11}\over 4}
\bigg]\Bigg\}
\cr
&~~~~~~~~~~~~~~~~~~~~~~~~~~~~~~~~~~~~~~~~~
\cdot~exp\left[{{24K^{3\over 2}}\over V}\left(1-\left(1-{{a^2V}\over {6K}}
\right)^{3\over 2}\right)\right]
\cr
&+~{{2i}\over {a^3}}\Bigg\{-{5\over {24}}\left({{aV}\over {24K}}\right)^3
\left[1-{{a^2V}\over {6K}}\right]^{-{{13}\over 4}}+{5\over {K^{3\over 2}}}
\left({V\over {144}}\right)^2 \left[1-{{a^2V}\over {6K}}\right]^{-{7\over 4}}
\cr
&~~~~+{V\over {48}}\left(2-{{a^2V}\over {6K}}\right)\left[1-{{a^2V}\over {6K}}
\right]^{-{5\over 4}}-K^{3\over 2}\left({{a^2V}\over {6K}}+2\right)\left[1-
{{a^2V}\over {6K}}\right]^{1\over 4}\Bigg\}
\cr
&~~~~~~~~~~~~~~~~~~~~~~~~~~~~~~~~~~~~~~\cdot~~exp\left[{{24K^{3\over 2}}\over 
V}\left(1+\left(1-{{a^2V}\over {6K}}\right)^{3\over 2}\right)\right]
\cr
&~+\Bigg\{2i{{V^\prime}\over {a^5}}
\left({{7K}\over 4}-{{35}\over {576a^4}}\right)
\biggl[{{a^2}\over {24K}}\left[1-{{a^2V}\over {6K}}\right]^{-{5\over 4}}-
{{24K^{3\over 2}}\over {V^2}}\left[1-{{a^2}\over {6K}}\right]^{-{1\over 4}}
\cr
&~~~-6{{a^2K^{1\over 2}}\over V}\left[1-{{a^2V}\over V}\right]^{1\over 4}
-{{24K^{3\over 2}}\over {V^2}}\left[1-{{a^2V}\over {6K}}
\right]^{5\over 4}\biggr]+i{{V^\prime}\over {12a^8}}
\biggl[{5\over {288}}{{a^3V}\over {K^2}}\left[1-{{a^2V}\over {6K}}
\right]^{-{9\over 4}}
\cr
&~~~+\left({a\over {12K}}-2{{aK^{1\over 2}}\over V}
\right)\left[1-{{a^2V}\over {6K}}\right]^{-{5\over 4}}
+\left(288{{aK^2}\over {V^2}}-2{{aK^{1\over 2}}\over V}\right)
\left[1-{{a^2V}\over {6K}}\right]^{1\over 4}
\cr
&~~~+72{{a^3K}\over V}\left[1-{{a^2V}\over {6K}}\right]^{3\over 4}
+288{{aK^2}\over {V^2}}\left[1-{{a^2V}\over {6K}}\right]^{7\over 4}
\biggr]
-i{{V^\prime}\over {32a^7}}\bigg[{5\over {384}}{{a^4V^2}\over {K^3}}
\left[1-{{a^2V}\over {6K}}\right]^{-{{13}\over 4}}
\cr}
$$
$$\eqalign{&~~~~+\left({{25}\over {288}}{{a^2V}\over {K^2}}
-{5\over 6}{{a^2}\over {K^{1\over 2}}}\right)
\left[1-{{a^2V}\over {6K}}\right]^{-{9\over 4}}
-{5\over {24}}{{a^4V}\over {K^{3\over 2}}}\left[1-{{a^2V}\over {6K}}
\right]^{-{7\over 4}}
\cr
&~~~~+\left({1\over {12K}}-2{{K^{1\over 2}}\over V}\right)
\left[1-{{a^2V}\over {6K}}\right]^{-{5\over 4}}
-{5\over 6}{{a^2}\over {K^{1\over 2}}}
\left[1-{{a^2V}\over {6K}}\right]^{-{3\over 4}}    
-18a^4\left[1-{{a^2V}\over {6K}}\right]^{-{1\over 4}}
\cr
&~~~+\left(288{{K^2}\over {V^2}}-2{{K^{1\over 2}}\over V}\right)
\left[1-{{a^2V}\over {6K}}\right]^{1\over 4}
+\left(72{{a^2K}\over V}-3456{{a^2K^{5\over 2}}\over {V^2}}\right)
\cdot\left[1-{{a^2V}\over {6K}}\right]^{3\over 4}
\cr
&~~~-864{{a^4K^{3\over 2}}\over V}\left[1-{{a^2V}\over {6K}}\right]^{5\over 4}
+288{{K^2}\over {V^2}}\left[1-{{a^2V}\over {6K}}\right]^{7\over 4}
-3456{{a^2K^{5\over 2}}\over {V^2}}\left[1-{{a^2V}\over {6K}}
\right]^{9\over 4}\bigg]
\cr
&+i{{V^\prime}\over {432a^6}}
\biggl[{{65}\over {4608}}{{a^5V^3}\over {K^4}}
\left[1-{{a^2V}\over {6K}}\right]^{-{{17}\over 4}}
+\left({{15}\over {128}}{{a^3V^2}\over {K^3}}-{5\over 8}{{a^3V}\over 
{K^{1\over 2}}}\right)\left[1-{{a^2V}\over {6K}}\right]^{-{{13}\over 4}}
\cr
&~~~~+{5\over {18}}{{a^5V^2}\over {K^{5\over 2}}}
\left[1-{{a^2V}\over {6K}}\right]^{-{{11}\over 4}}
+\left({5\over {24}}{{aV}\over {K^2}}-{5\over 2}{a\over {K^{1\over 2}}}
\right)\left[1-{{a^2V}\over {6K}}\right]^{-{9\over 4}}
+\left(10a^3-{{25}\over {12}}{{a^3V}\over {K^{3\over 2}}}\right)
\cr
&~~~~\cdot\left[1-{{a^2V}\over {6K}}\right]^{-{7\over 4}}
+{{a^5V}\over K}\left[1-{{a^2V}\over {6K}}\right]^{-{5\over 4}}
-{5\over 2}{a\over {K^{1\over 2}}}\left[1-{{a^2V}\over {6K}}
\right]^{-{3\over 4}}
\cr
&~~~~+\left(864{{a^3K^{3\over 2}}\over V}-80a^3\right)
\left[1-{{a^2V}\over {6K}}\right]^{-{1\over 4}}
+576a^5K^{1\over 2}\left[1-{{a^2V}\over {6K}}\right]^{1\over 4}
\cr
&~~~~-10368{{aK^{5\over 2}}\over V}
\left[1-{{a^2V}\over {6K}}\right]^{3\over 4}
+\left(41472{{a^3K^3}\over {V^2}}-1728{{a^3K^{3\over 2}}\over V}\right)
\left[1-{{a^2V}\over {6K}}\right]^{5\over 4}
\cr
&~~~~+10368{{a^5K^2}\over V}\left[1-{{a^2V}\over {6K}}\right]^{7\over 4}
-10368{{aK^{5\over 2}}\over {V^2}}\left[1-{{a^2V}\over {6K}}\right]^{9\over 4}
+41472{{a^3K^3}\over {V^2}}\left[1-{{a^2V}\over {6K}}
\right]^{{11}\over 4}\biggr]\Bigg\}
\cr
&~~~~~~~~~~~~~~~~~~~~~~~~~~~~~~~~~~~~~~~~~~~~~~\cdot~
exp\left[{{24K^{3\over 2}}\over V}\left(1+\left(1-{{a^2V}\over {6K}}
\right)^{3\over 2}\right)\right]
\cr}
\eqno(23)
$$

Terms containing $V^\prime(\phi)$ are negligible if 
$\vert V^{-1}V^\prime(\phi)\vert << 1$ in Planck units, which holds
for scalar potentials in standard inflationary models.  In
string cosmology, slow-roll inflation occurs because of the
flatness of the potential in the chiral field directions [42], but it
can be verified that the terms involving the derivative with respect to
the dilaton, $V^\prime(\Phi)$, are comparable in magnitude to the
other terms in $H_1 \Psi_0$ when $V(\Phi)$ does not contain an effective
cosmological constant in addition to expressions multiplied by powers of 
of ${{e^{-\Phi}}\over {g_4^2}}$.

Nevertheless, in the parameter range where the derivative with respect 
to $\Phi$ can be ignored, the general solution to the equation
$H_0 \Psi_1= -H_1\Psi_0$, given two linearly independent solutions of 
the homogeneous equation,
$\left({1\over {24}}{d\over {da}}{1\over a}{d\over {d a}}~-~6aK
~+~a^3V(\Phi)\right)\Psi_0~\simeq~0$, would be   
$$\Psi_1~\approx~C_1~\Psi_{0_1}~+~C_2~\Psi_{0_2}~-~24~\Psi_{0_2}
~\int~\Psi_{0_1}~{{H_1 \Psi_0}\over W}~a~da
           ~+~24~\Psi_{0_1}~\int~\Psi_{0_2}~{{H_1\Psi_0}\over W}~a~da
\eqno(24)
$$
where $W~=~\Psi_{0_1}{d\over {da}}\Psi_{0_2}~-~\Psi_{0_2}~{d\over {da}}
\Psi_{0_1}$.  Substituting $H_1\Psi_{0_{NB}}$ or $H_1\Psi_{0_T}$ into this 
formula and imposing the appropriate boundary conditions on $\Psi_1$ 
provides the correction of order ${{e^{-\Phi}}\over {g_4^2}}~\Psi_1$ to 
the standard wave function $\Psi_0$.

Amongst the solutions to the equations of motion of a one-loop effective
Lagrangian, having the same form as ${\cal L}_{eff.}$ in equation (1) and 
containing only the dilaton field, are the homogeneous and isotropic 
solutions that begin with a Gauss-Bonnet phase 
$a(t) \sim e^{-{{\omega_1}\over t}}$ and continue 
to the FRW phase $a(t) \sim t^{1\over 3}$ [43].  The stability of these 
solutions with respect to linearized tensor perturbations [44] 
is determined by the effective adiabatic index 
$\Gamma = {2\over 3} \left(1 -{{a {\ddot a}}\over {{\dot a}^2}}
\right)$ [45].  The Gauss-Bonnet phase is unstable since $\Gamma < {5\over 3}$,
and the background geometry is eventually described by a stable FRW metric
with $\Gamma = 2$.

Stability of the geometry, against perturbations of the metric that 
do not change the topology, follows from the
positive-energy theorem, which is applicable when the metric admits
Killing spinors.  The requirement of covariance and spatial 
homogeneity implies the following form for the Killing spinor equation:
$$\nabla_\mu~\eta~+~i{{k(t)}\over 2}\gamma_\mu~\eta~=~0
\eqno(25)
$$
for some function $k(t)$.  Considering the spatial components 
of equation (25) and applying the 
\hfil\break
commutator of the modified 
covariant derivative to the spinor $\eta$, the integrability relation [46] is 
${1\over 4}R_{ij}{}^{\alpha\beta}\gamma_{\alpha\beta}\eta
={{k^2(t)}\over 2}\gamma_{ij}\eta$, 
implying that the four-dimensional space-time should be
foliated by three-dimensional spaces satisfying the condition 
$R_{ijkl}={{k^2(t)}\over 2}(g_{ik}g_{jl}-g_{il}g_{jk})$
and $k^2(t)=a^{-2}(K+{\dot a}^2)$.

The time-space component of the commutator condition is
$$\left({1\over 2}R_{0i}{}^{0j}\gamma_{0j}~+~{1\over 4}R_{0i}{}^{jk}
\gamma_{jk}\right)\eta~=~-{i\over 2}{\dot k(t)}\gamma_i \eta
~+~{{k^2(t)}\over 2}\gamma_{0i}\eta
\eqno(26)
$$ 
which implies that
$${1\over 2}{{\ddot a(t)}\over {a(t)}}~\gamma_{0i}\eta~=~
{{k^2(t)}\over 2}\gamma_{0i}\eta~-~{i\over 2}{\dot k(t)}\gamma_i \eta
\eqno(27)
$$
Using the Dirac representation of the gamma matrices in four dimensions,
\hfil\break
$\gamma_0=\left(\matrix{I& 0
                          \cr
                          0& -I
                           \cr}
                               \right)$,
and setting $\eta$ equal to $c_1 \left({{\eta_1}\atop 0}\right)+
c_2\left({0\atop {\eta_2}}\right)$, it follows that

$$c_1\left({{{\ddot a}(t)}\over {a(t)}}~-~k^2(t)~+~i{\dot k(t)}\right)
\left({{\eta_1}\atop 0}\right)
~+~c_2 \left(-{{{\ddot a}(t)}\over {a(t)}}~+~k^2(t)~+~i{\dot k(t)}\right)
\left({0\atop {\eta_2}}\right)~=~0
\eqno(28)
$$

There are three types of solutions:  
\vskip 3pt
\noindent
(i)~$c_1,~c_2\ne~0$,~${\dot k(t)}~=~0$,~ 
${{{\ddot a}(t)}\over {a(t)}}~-~k^2(t)~=~0$,
\hfil\break
(ii)~$c_2~=~0,~{ {{\ddot a}(t)}\over {a(t)}}~-~k^2(t)~+~i{\dot k(t)}~=~0$, 
\hfil\break
(iii)~$c_1~=~0,~-{ {{\ddot a}(t)}\over {a(t)}}~+~k^2(t)~+~i{\dot k(t)}~=~0$.

None of these conditions are satisfied by the scale factors
$a(t) \sim e^{-{{\omega_1}\over t}}$ or $a(t) \sim t^{1\over 3}$.
This implies that the positive-energy theorem is not applicable to these
background geometries, allowing for the possibility that they are
unstable at a non-perturbative level.  Moreover, these classical solutions 
may represent only local extrema of the action, providing sub-dominant 
contributions to the string path integral, since the absence of stability 
at the non-perturbative level suggests that there is tunneling 
from these solutions to more stable background geometries.  

The Majorana condition implies that $c_1,~c_2~\ne~0$ and the Killing spinor
belongs to first category.  When $K=1$, both constraints are satisfied 
by the scale factor $a(t) = a_0~cosh({t\over {a_0}}),~k(t)={1\over {a_0}}$,  
representing a cosmological bounce solution.  
When $K=0$, the Killing spinor conditions require $a(t)=a_0~e^{\lambda t}$,
and if $K=-1$, they imply that $a(t)=a_0~sinh({t\over {a_0}})$ or
$a(t)=a_0~sin({t\over {a_0}})$.  

Since the positive-energy theorem is applicable to these space-times, 
they will be stable at the non-perturbative level within the class of metrics 
representing the same topology.  It is of interest to note that the $K=0$ and 
$K=1$ solutions do not have an initial singularity and that they 
provide adequate classical cosmological models through the inflationary 
epoch, as the scale factors $a(t)$ increase exponentially with time.

It is customary to attribute physical significance to a wave function
only when there is an appropriate classical limit.  While the metric
with an initial Gauss-Bonnet phase and a perturbatively stable FRW phase
represents a cosmological model which is realistic at later times, the
stable, non-singular, exponentially expanding space-times are
preferable as background geometries in this quantum cosmological model,
because they describe the inflationary epoch, where quantum effects
are still relevant.

The Euler-Lagrange equation of motion for $\Phi$ is
$${\ddot \Phi}~+~6{{e^{-\Phi}}\over {g_4^2}}\left[{{{\ddot a}
(K+{\dot a}^2)}\over {a^3}}~-~{{{\dot a}^2({\dot a}^2+3K)}\over {a^4}}\right]
~+~V^\prime(\Phi)~=~0
\eqno(29)
$$
With the scale factor $a(t)=a_0~cosh \left({t\over {a_0}}\right)$, and the 
heterotic string potential (13), this equation becomes
$$\eqalign{{\ddot \Phi}&~+~{6\over {a_0^4}}{{e^{-\Phi}}\over {g_4^2}}
~sech^2\left({t\over {a_0}}\right)
\left(1-2~tanh^2\left({t\over {a_0}}\right)\right)
~-~{{e^{-3\sigma_0}}\over {16}}{{e^{-\Phi}}\over {g_4^2}}
{d\over {dS}}\biggl\{{1\over S}~\vert {\cal W}(S)\vert^2
\cr
&~~~~~~~~~~~~~~~~~~~~~\times \left[{1\over S}-2{{{\cal W}^\prime(S)}\over 
{{\cal W}(S)}}\right]^2
\left[-{1\over {S^2}}+
2{{{\cal W}^\prime(S)^2-{\cal W}(S){\cal W}^{\prime\prime}(S)}\over 
{{\cal W}(S)^2}}\right]^{-1}\biggr\}=0
\cr}
\eqno(30)
$$
Using the superpotential ${\cal W}(S)=c+h \left(1+{{3S}\over {b_0}}\right)
e^{-{{3S}\over {2b_0}}}$ [47], the equation of motion becomes 

$$\eqalign{&{\ddot\Phi}+{6\over {a_0^4}}{{e^{-\Phi}}\over {g_4^2}}
sech^2\left({t\over {a_0}}\right)\left(1-2~tanh^2
\left({t\over {a_0}}\right)\right)-{{e^{-3\sigma_0}}\over {16}}
{{e^{-\Phi}}\over {g_4^2}}\Biggl\{\biggl[-{1\over {S^2}}\biggl[(c+h)+
\cr
&~~~~~~~{{3hS}\over {b_0}}e^{-{{3S}\over {2b_0}}}\biggr]^2
+{1\over S}\left[-{{18hcS}\over {b_0^2}}e^{-{{3S}\over {2b_0}}}
+{{3h^2}\over {b_0}}\left(1+{{6S}\over {b_0}}e^{-{{3S}\over {2b_0}}}
\right)\right]\biggr]
\cdot\left[{1\over S}-2{{{\cal W}^\prime(S)}\over {{\cal W}(S)}}\right]^2
\cr
&~~~~~~~~~~~~~~~~~~~~~~~~~~~\cdot\left[-{1\over {S^2}}+
2{{{\cal W}^\prime(S)^2-{\cal W}(S)
{\cal W}^{\prime\prime}(S)}\over {{\cal W}(S)^2}}\right]^{-1}
+{1\over S}\left[(c+h)+{{3hS}\over {b_0}}
e^{-{{3S}\over {2b_0}}}\right]^2
\cr
&~~~~~~~~~~\cdot\biggl[2\left({1\over S}-2{{{\cal W}^\prime(S)}\over 
{{\cal W}(S)}}\right)+2\left({1\over S}-2{{{\cal W}^\prime(S)}\over 
{{\cal W}(S)}}\right)^2
\cdot \biggl({1\over {S^2}}+2{{{\cal W}(S){\cal W}^{\prime\prime}(S)
-{\cal W}^\prime(S)^2}\over {{\cal W}(S)^2}}\biggr)^{-2}
\cr
&~~~~~~~~~~~~~~~~~~~~~~~~~~~~~~~~~~~~~~~\biggl(-{1\over {S^3}}
+{ {{\cal W}^{\prime\prime\prime}(S)}\over {{\cal W}(S)}}
-3{{{\cal W}^\prime(S){\cal W}^{\prime\prime}(S)}\over {{\cal W}(S)^2}}
+2{{{\cal W}^\prime(S)^3}\over {{\cal W}(S)^3}}\biggr)\biggr]
\Biggr\}~=~0
\cr}
\eqno(31)
$$
which reduces to 

$${\ddot \Phi}~+~
{6\over {a_0^4}}sech^2\left({t\over {a_0}}\right)\left(1-2~tanh^2
\left({t\over {a_0}}\right)\right)
{{e^{-\Phi}}\over {g_4^2}}
~-~{{27}\over 8}{{h^2e^{-3\sigma_0}}\over {b_0^3}}~{{e^{-2\Phi}}\over {g_4^4}}
~+~O(e^{-3\Phi})~=~0
\eqno(32)
$$
when $c$ is set equal to $-h$.  Based on the leading-order term in 
$V^\prime(\Phi)$, the dilaton field at large times is approximately 

$$\eqalign{\Phi(t)~&\simeq~ln~\left\vert~
{3\over {2a_0^2g_4^2}}\left(1+{\sqrt{1+{3\over 8}{{a_0^2h^2
e^{-3\sigma_0}}\over {b_0^3}} }}\right)
~cosh({\sqrt C}t)\right\vert
\cr
C~&=~{4\over {a_0^2}}
\cr}
\eqno(33)
$$
and $\Phi(t)~\to~{\sqrt C}t~+~ln~\left[~{3\over {4a_0^2g_4^2}}
\left(1+{\sqrt{1+{3\over 8}{{a_0^2h^2e^{-3\sigma_0}}\over {b_0^3}} }}\right)
\right]~\equiv~{\sqrt C}t~+~D$ as $t~\to~\infty$, representing the linear 
dilaton solution [48][49].  

The equation of motion for $a(t)$ is
$$\eqalign{30{\dot a}^2~-~6K~+~&6{{e^{-\Phi}}\over {g_4^2}}({\ddot \Phi}
~-~{\dot \Phi}^2)({\dot a}^2~+~K)
~-~12\left(a~-~{{e^{-\Phi}}\over {g_4^2}}{\dot \Phi}{\dot a}
\right){\ddot a}
\cr
~&-~18{{e^{-\Phi}}\over {g_4^2}}{{\dot a}\over a}
{\dot \Phi}({\dot a}^2~+~K)~-~{3\over 2}a^2{\dot \Phi}^2~+~3a^2 V(\Phi)~=~0
\cr}
\eqno(34)
$$

Given the string potential
$$V(\Phi)~=~{{g_4^2 e^\Phi e^{-3\sigma_0}}\over {16}}
{ {\left[(c+h)+{{3h}\over {b_0}}{{e^{-\Phi}}\over {g_4^2}}
e^{-3{{e^{-\Phi}}\over {2b_0g_4^2}} }\right]^2~\cdot~
\left[\left(c+h+{{9h}\over {b_0^2}}{{e^{-2\Phi}}\over {g_4^4}}\right)
e^{-3{{e^{-\Phi}}\over {2b_0g_4^2}}}\right]^2}
\over
{\left[(c+h)^2+{{3h(h+c)}\over {b_0}}{{e^{-\Phi}}\over {g_4^2}}
-{1\over 4}{{h(50h+81c)}\over {b_0^2}}{{e^{-2\Phi}}\over {g_4^4}}\right]}}
\eqno(35)
$$
it is convenient to set $(c+h)=ke^{-{{\sqrt C}\over 2}t}$ instead of 
equating it to zero.

When $K=1$ and $a(t)=a_0~cosh \left({t\over {a_0}}\right)$, to leading order,
equation (34) implies $18-{3\over 2}a_0^2 C + {3\over {16}}a_0^2g_4^2 
e^{-3\sigma_0}k^2 e^D~=~0$.  While there is a positive solution for $C$ 
when $k$ is real, it does not equal ${4\over {a_0^2}}$.  A similar result
occurs when $K=-1$.  If $K=0$ and $a(t)=a_0~e^{\lambda t}$, there is again a 
linear dilaton solution $\Phi(t)~\sim~ln~\left\vert {{27}\over {4C}}
{{h^2e^{-3\sigma_0}}\over {b_0^3g_4^4}}cosh({\sqrt C}(t-t_0))\right\vert$. 
To leading order, the equation of motion for $a(t)$ gives rise to the relation
$18\lambda^2-{3\over 2}C+{3\over {16}}g_4^2 e^{-3\sigma_0}k^2 e^D=0$
which yields a positive value of $C$ that can be adjusted with $\lambda$.

Since a linear growth for $\Phi$ and an exponential scale factor $a(t)$,
representing a stable classical background geometry, have been obtained from 
an approximate solution to the equations of motion for the
action (4) derived from heterotic string theory, these properties 
should characterize configurations that are relatively more probable in the 
minisuperspace $\{a(t),\Phi(t)\}$.  This allows one to compare directly 
predictions of the quantum cosmological model with standard inflationary 
cosmology.  Moreover, supersymmetry ensures that the FRW 
universes with scale factors $a_0~cosh\left({t\over {a_0}}\right)$, 
$a_0~e^{\lambda t}$ and $a_0~sinh\left({t\over {a_0}}\right)$ are stable,
so that inflation occurring in such exponentially expanding space-times
would be terminated only when this symmetry is broken.  Other scale factors, 
consistent with astrophysical observations, can be introduced consistently 
within this model for later epochs, once supersymmetry is broken.

The same techniques can be applied to other superstring effective actions
with higher-order curvature terms, and again, the conjugate momenta and the 
Hamiltonian may be derived, with the solution to the Wheeler-DeWitt equation
satisfying specified boundary conditions.  Since the calculation of the 
wave function is equivalent to the evaluation of the path-integral 
over 4-metrics between initial and final times, the conclusions are 
consistent with the space-time foam picture, and theoretical expectation 
values of functions of the minisuperspace coordinates can be compared to
observations of these cosmological variables.
\vskip 10pt
\centerline{\bf Acknowledgements}
\noindent
Research on this project has been partly supported by an ARC Small Grant.
\vfill\eject
\centerline{\bf References}
\vskip 10pt
\parskip=5pt
\baselineskip=3pt
\item{[1]} S. W. Hawking and J. B. Hartle, Phys. Rev. ${\underline{D28}}$
(1983) 2960-2975
\item{[2]} K. Stelle, Phys. Rev. ${\underline{D16}}$ (1977) 953-969
\item{[3]} L. Querella, ${\underline{Variational~Principles~and~Cosmological
~Principles~in}}$
\hfil\break
${\underline{Higher-Order~Gravity}}$, Doctoral dissertation - Universit{\`e}
de Liege (1998) 
\item{[4]} I. Antoniadis, J. Rizos and K. Tamvakis, Nucl. Phys.
${\underline{B415}}$ (1994) 497-514
\item{[5]} P. Kanti, J. Rizos and K. Tamvakis, Phys. Rev. ${\underline{D59}}$
(1999) 083512: 1-12
\item{[6]} D. J. Gross and J. H. Sloan, Nucl. Phys. ${\underline{B291}}$
(1987) 41-89
\item{[7]} P. W. Higgs, Nuovo Cimento ${\underline{11}}$ (1959) 816-820
\item{[8]} G. V. Bicknell, J. Phys. ${\underline{A7}}$ (1974) 341-345
\item{[9]} B. Whitt, Phys. Lett. ${\underline{B145}}$ (1984) 176-178
\item{[10]} H.-J. Schmidt, Astron. Nachr. ${\underline{308}}$ (1987) 183-188
\item{[11]} V. F. Mukhanov, L. A. Kofman and D. Yu. Pogosyan, 
Phys. Lett. ${\underline{B193}}$ (1987) 427-432
\item{[12]} J. D. Barrow and S. Cotsakis, Phys. Lett. ${\underline{B214}}$
(1988) 515-518
\item{[13]} K. Maeda, Phys. Rev. ${\underline{D39}}$ (1989) 3159-3162
\item{[14]} G. Magnano, M. Ferraris and M. Francaviglia, Gen. Rel.
Grav. ${\underline{19}}$ (1987) 465-479
\item{[15]} A. Jakubiec and J. Kijowski, Phys. Rev. ${\underline{D37}}$
(1988) 1406-1409
\item{[16]} M. Ferraris, M. Francaviglia and G. Magnano, Class. Quantum
Grav. ${\underline{5}}$ (1988) L95
\item{[17]} H. J. de Vega and N. Sanchez, `Lectures on String Theory in 
Curved Space-Times', in 
\hfil\break
${\underline{String~Gravity~and~Physics~at~the~
Planck~Energy~Scale}}$ (Erice, 1995) 11-63
\item{[18]} A. L. Maroto and I. L. Shapiro, Phys. Lett. ${\underline{B414}}$
(1997) 34-44
\item{[19]} M. B. Mijic, M. S. Morris, W.- M. Suen, Phys. Rev. 
${\underline{D34}}$ (1986) 2934-2946
\item{[20]} A. Berkin, Phys. Rev. D ${\underline{D44}}$ (1991) 1020-1027
\item{[21]} S. W. Hawking and J. C. Luttrell, Phys. Lett. ${\underline{B143}}$
(1984) 83-86
\item{[22]} A. Dobado and A. L. Maroto, Phys. Rev. ${\underline{D52}}$
(1995) 1895-1901
\item{[23]} F. D. Mazzitelli, Phys. Rev. ${\underline{D45}}$ (1992)
2814-2823
\item{[24]} M. D{\"u}tsch and B. Schroer, `Massive Vectormesons 
and Gauge Theory', hep-th/9906089
\item{[25]} S. W. Hawking, `Quantum Cosmology', ${\underline{Relativity,~
Groups~and~Topology~II}}$, 
\hfil\break
Les Houches 1983, Session XL, edited by
B. S. De Witt and R. Stora (North-Holland, Amsterdam, 1984) 333-379  
\item{[26]} A. Vilenkin, Phys. Rev. ${\underline{D32}}$ (1985) 2511-2548
\item{[27]} A. D. Linde ${\underline{Inflation~and~Quantum~Cosmology}}$, ed. 
by R. H. Brandenburger 
\hfil\break
(Boston: Academic Press, 1990)
\item{[28]} N. A. Lemos, Phys. Rev. ${\underline{D53}}$ (1996) 4275-4279
\item{[29]} M. D. Pollock, Nucl. Phys. ${\underline{B315}}$ (1989) 
528-540
\item{[30]} M. D. Pollock, Nucl. Phys. ${\underline{B324}}$ (1989) 187-204
\item{[31]} M. D. Pollock, Int. J. Mod. Phys. A ${\underline{7}}$(17)
(1992) 4149-4165
\item{[32]} M. D. Pollock, Int. J. Mod. Phys. D ${\underline{4}}$(3) (1995) 
305-326
\item{[33]} S. Davis, `Higher-Derivative Quantum Cosmology', Proceedings
of the ACGRG2 Meeting, 6 - 10 July 1998, Gen. Rel. Grav. ${\underline{32}}$
(3) (2000) 541-551
\item{[34]} N. Kontoleon and D. L. Wiltshire, Phys. Rev. ${\underline{D59}}$
(1999) 063513: 1-8
\item{[35]} S. Ferrara, C. Kounnas and M. Porrati, Phys. Lett.
${\underline{B181}}$ (1986) 263-268
\item{[36]} P. Binetruy, S. Dawson, I. Hinchcliffe and M. K. Gaillard, 
Phys. Lett. ${\underline{B192}}$ (1987) 377-384
\item{[37]} I. Antoniadis and T. R. Taylor, Nucl. Phys. (Proc. Suppl.)
${\underline{41}}$ (1995) 279-287
\item{[38]} D. Bailin and A. Love, Phys. Rep. ${\underline{315}}$(4-5)
(1999) 285-408
\item{[39]} A. Vilenkin, Phys. Rev. ${\underline{D30}}$ (1984) 509 - 511
\item{[40]} A. Vilenkin, Phys. Rev. ${\underline{D33}}$ (1986) 3560-3569
\item{[41]} A. Vilenkin, Phys. Rev. ${\underline{D37}}$ (1988) 888-897 
\item{[42]} J. A. Casas, G. B. Gelmini, A. Riotto, Phys. Lett.
${\underline{B459}}$ (1999) 91 - 96
\item{[43]} J. Soda, M. Sakagami and S. Kawai, `Novel Instability in
Superstring Cosmology',
\hfil\break
gr-qc/9807056; to be published in the proceedings of International Seminar 
on Mathematical Cosmology (ISMC 98) Potsdam, Germany 30 Mar - 4 Apr 1998
\item{[44]} S. Kawai, M. Sakagami and J. Soda, Phys. Lett. 
${\underline{B437}}$ (1998) 284-290
\item{[45]} M. B. Green, J. H. Schwarz and E. Witten, 
${\underline{Superstring~Theory:~Volume~2}}$ (Cambridge: 
Cambridge University Press, 1987)
\item{[46]} P. van Nieuwenhuizen and N. Warner, Commun. Math. Phys.
${\underline{93}}$ (1984) 277-284 
\item{[47]} P. Binetruy, S. Dawson, I. Hinchliffe and M. K. Gaillard,
Phys. Lett. ${\underline{B192}}$ (1987) 377 - 382
\item{[48]} M. Gasperini, `Birth of the universe in string cosmology',
${\underline{Fourth~Paris~Cosmology}}$
\hfil\break
$\underline{Colloquium}$, eds. H. J. De Vega
and N. Sanchez (Singapore: World Scientific, 1998), p.85
\item{[49]} D. Clancy, A. Feinstein, J. E. Lidsey and R. Tavakol,
Phys. Rev. ${\underline{D60}}$ (1999) 043503:1-10

\end